# Probing ultrafast ππ*/nπ* internal conversion in organic chromophores via K-edge resonant absorption

**Authors:** T. J. A. Wolf[1], R. H. Myhre[1,2], J. P. Cryan[1], S. Coriani[3,4], R. J. Squibb[5], A. Battistoni[1], N. Berrah[6], C. Bostedt[7,8,9], P. Bucksbaum[1,10], G. Coslovich[7], R. Feifel[5], K. J. Gaffney[1], J. Grilj[11], T. J. Martinez[1,12], S. Miyabe[1,12,13], S. P. Moeller[7], M. Mucke[14], A. Natan[1], R. Obaid[6], T. Osipov[7], O. Plekan[15], S. Wang[1], H. Koch[1,2]*, M. Gühr[1,16]*

[1]Stanford PULSE Institute, SLAC National Accelerator Laboratory, Menlo Park, CA 94025, USA.
[2]Department of Chemistry, Norwegian University of Science and Technology, NO-7491 Trondheim, Norway.
[3]Dipartimento di Scienze Chimiche e Farmaceutiche, Università degli Studi di Trieste, Italy.
[4]Aarhus Institute of Advanced Studies, Aarhus University, DK-8000 Aarhus C, Denmark.
[5]Department of Physics, University of Gothenburg, Gothenburg, Sweden.
[6]Department of Physics, University of Connecticut, Storrs, Connecticut 06269, USA.
[7]Linac Coherent Light Source, SLAC National Accelerator Laboratory, Menlo Park, California 94720, USA.
[8]Argonne National Laboratory, Lemont, Illinois 60439, USA.
[9]Department of Physics and Astronomy, Northwestern University, Evanston, Illinois 60208, USA.
[10]Department of Physics, Stanford University, Stanford, CA 94305, USA.
[11]Laboratory of Ultrafast Spectroscopy, Ecole Polytechnique Federal de Lausanne, CH 1015 Switzerland.
[12]Department of Chemistry, Stanford University, Stanford, California 94305, USA.
[13]Laser Technology Laboratory, RIKEN, Wako, Saitama 351-0198, Japan
[14]Department of Physics and Astronomy, Uppsala University, Box 516, SE-751 20 Uppsala, Sweden.
[15]Elettra-Sincrotrone Trieste, I-34149 Basovizza, Trieste, Italy.
[16]Institut für Physik und Astronomie, Universität Potsdam, 14476 Potsdam, Germany.

*Correspondence to: mguehr@uni-potsdam.de, henrik.koch@ntnu.no

*Abstract:* Organic chromophores with heteroatoms possess an important excited state relaxation channel from an optically allowed ππ* to a dark nπ*state. We exploit the element and site specificity of soft x-ray absorption spectroscopy to selectively follow the electronic change during the ππ*/nπ* internal conversion. As a hole forms in the n orbital during ππ*/nπ* internal conversion, the near edge x-ray absorption fine structure (NEXAFS) spectrum at the heteroatom K-edge exhibits an additional resonance. We demonstrate the concept with the nucleobase thymine, a prototypical heteroatomic chromophore. With the help of time resolved NEXAFS spectroscopy at the oxygen K-edge, we unambiguously show that  ππ*/nπ* internal conversion takes place within (60±30) fs. High-level coupled cluster calculations on the isolated molecules used in the experiment confirm the superb electronic structure sensitivity of this new method for excited state investigations.

The efficient conversion of light into other forms of energy plays a key role in many processes such as photosynthesis or human vision (*1*, *2*). It is well established that the efficiency of these processes is facilitated by coupled ultrafast electronic and nuclear dynamics that cannot be described using the Born-Oppenheimer approximation (BOA). The breakdown of the BOA implies that the fundamental details of such mechanisms are notoriously difficult to understand: they proceed on an ultrafast timescale and occur mostly at positions where potential energy surfaces come close or even intersect. From an experimental point of view, it is highly desirable to access the nuclear geometry as well as the electronic degrees of freedom separately, in order to compare them to quantum simulations. The transient nuclear geometry can be best studied by time resolved diffraction techniques using short x-ray (*3*) or electron pulses (*4*). X-ray spectroscopy methods provide high electronic state selectivity, as shown prominently for metal containing molecules at their K- and L-edges (*5–7*). We demonstrate the ability of time-resolved soft x-ray spectroscopy to differentiate between distinct electronic excited states of organic molecules, closing an important gap in the classes of molecules that can be investigated.

In this article, we concentrate on the internal conversion between excited states of different electronic character, which is a crucial path for photoenergy conversion in organic molecules. Organic chromophores exhibit strongly absorbing ππ* excited states, which can be described in a single electron Hartree-Fock (HF) picture as an electron-hole pair in a formerly occupied and an unoccupied molecular orbital (MO), both with π symmetry. Many of these chromophores, such as azo-switches (*3*, *8*, *9*), nucleobases (*10–13*), and amino acids (*14*), also contain heteroatoms with electron lone pairs. They therefore exhibit nπ* excited states, with a hole in a heteroatom-centered lone pair (n) orbital and an electron in a π* orbital. The ππ*/nπ* internal conversion provides photochemical pathways for reactions like cis-trans isomerization and intersystem crossing to the triplet manifold of electronic states governed by the El Sayed selection rules (*15*). Unlike ππ* excited states, nπ* states are usually not directly accessible due to low absorption cross-sections from the ground state. The ππ*/nπ* internal conversion through conical intersections is therefore a crucial gateway process for photochemistry.

The preferential localization of the n orbital at the heteroatom has wide-reaching implications for resonant core level spectroscopy using x-rays. In general, near edge x-ray absorption fine structure (NEXAFS) spectra show isolated features due to resonant states below the core ionization edge of an element. Those features are due to transitions from this element's core orbital to an unoccupied valence orbital, for instance a $\pi^*$ orbital. The core orbital is confined to the immediate vicinity of one particular atom (see Fig. 1A). The core-to-valence absorption cross-section is strongly dependent on the spatial overlap between the core and the empty valence orbital (*24*, *25*). Since core ionization potentials of carbon, nitrogen and oxygen are more than 100 eV apart, element- and site-specific probing of the local electronic structure in an organic molecule is possible with soft x-rays. In the case of excited states, the electron hole in a formerly occupied orbital enables an additional NEXAFS resonance. The spatial overlap makes a 1s-n transition from the strongly localized heteroatom 1s core level to the electron hole of an nπ* state more intense than the 1s-π transition to the delocalized π hole of a ππ* state. Therefore, we expect a negligibly weak time-resolved (TR) NEXAFS signature from the photoexcited ππ* state to transform into a strong nπ* state signature as the molecule undergoes ππ*/nπ* internal conversion, largely independent of geometry changes during the dynamics. For the isolated molecules used in this study, computationally demanding high level coupled cluster (CC) simulations (*16–18*) are now feasible and confirm the spectroscopic attribution based on orbital localization.

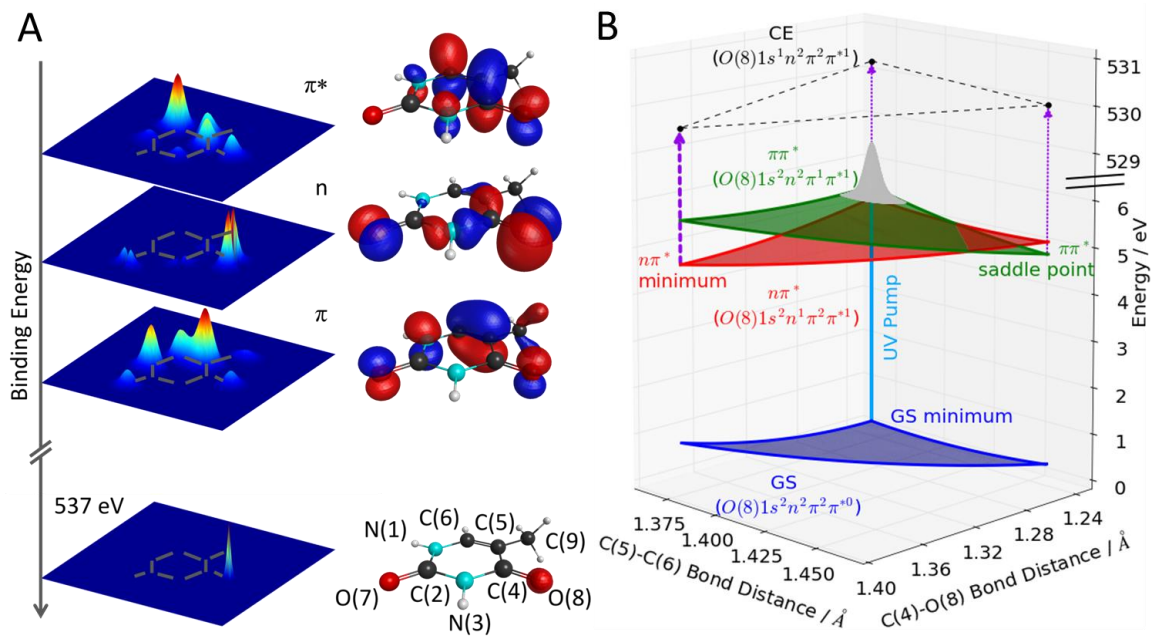

**Fig. 1**. *(A) Isosurface representations (right) and electron density projections onto the molecular plane for the three valence orbitals (Hartree-Fock/6-311G) involved in the characters of the two lowest lying excited states of thymine and a core orbital localized at oxygen(8). The electron density at the position of the core orbital differs strongly for the different valence orbitals. (B) Results from our coupled cluster investigation of the excited state topology along the two most relevant degrees of freedom for relaxation into the $n\pi^*$ minimum. All states are labeled with their electron configuration. Ultraviolet (UV) excitation of the ground state (GS) places a nuclear wavepacket (gray) on the $\pi\pi^*$ excited state. It relaxes through a conical intersection to a minimum in the $n\pi^*$ excited state. According to our calculations, only one core excited state (CE) characterized by an excitation at O(8) is relevant for interpretation of our near-edge x-ray absorption fine structure (NEXAFS) results.*

We test and exemplify the sensitivity of TR-NEXAFS spectroscopy on the $\pi\pi^*$/$n\pi^*$ transition in the benchmark chromophore thymine, since there is a rich literature on its excited states (see Refs. (*10–12, 19*) and citations therein). Thymine exhibits two high-lying occupied MOs, an oxygen localized n-orbital and a delocalized $\pi$-orbital (see Fig. 1A). Its lowest unoccupied MO ($\pi^*$) is similar to the $\pi$ MO in delocalization. The molecule can be excited at 267 nm to a $\pi\pi^*$ state; the lower-lying $n\pi^*$ state is optically dark.

Figure 1B shows a reduced potential energy sketch from the CC simulations along the two nuclear coordinates, which are expected to be most relevant for the molecular dynamics. In contrast to earlier theoretical studies (*20, 21*), the $\pi\pi^*$/$n\pi^*$ conical intersection seam is not isolated by a barrier from the Franck-Condon (FC) point, but directly accessible. After photoexcitation, the nuclear wavepacket is driven out of the FC region by a gradient along the C(5)-C(6) bond elongation towards a saddle point. On its way, it encounters the $\pi\pi^*$/$n\pi^*$ conical intersection seam, which indicates a rapid (on a 100 fs scale) internal conversion. In the nπ* excited state, a local minimum can be reached from the ππ*/nπ* conical intersection by O(8)-C(4) bond elongation (*22*).

The ππ* relaxation in thymine has been experimentally investigated using many methods available in ultrafast technology (*11*, *12*) including our own DUXAP study, where we investigated excited state nonresonant Auger spectra at the oxygen K-edge (*23*). It is challenging to attribute signals in ultrafast photoelectron, photoion, absorption, or DUXAP spectroscopy directly to a particular process, like internal conversion, since changes in both the electronic structure as well as the nuclear geometry influence the observables. In our own study we were e.g. preferentially sensitive to local C-O bond length changes during relaxation of the ππ* state. We demonstrate in the following, that our new TR-NEXAFS technique is strongly and selectively sensitive to the ultrafast ππ*/nπ* internal conversion.

The experimental ground state NEXAFS spectrum of thymine is shown in black in Fig. 2A. It exhibits a double peak π* resonance. Based on our CC calculations, and in agreement with earlier studies (*26*), we assign the lower energy peak at 531.4 eV to a linear combination of HF single electron excitations from the O(8) 1s orbital to several unoccupied π* orbitals with significant contributions from the aforementioned π* MO. The linear combination is such that the core excited state possesses a *high degree of electron localization at O(8)*, thus the strong absorption cross section in the Mbarn regime (see a discussion in the Supporting Material). The higher energy peak at 532.2 eV corresponds to an excitation from the O(7) 1s orbital to a different linear combination of *π\** MOs (*26*). The increase in intensity at photon energies beyond the π* resonances is predominantly due to a smooth feature of K-edge ionization at 537 eV and additional weak resonant transitions (*26*).

The NEXAFS spectrum taken 2 ps after UV excitation is shown in green in Fig. 2A. It is a superposition of the excited state spectrum and the ground state spectrum, which is weakened by transfer of an estimated 13 % of the population (see Supporting Material) to the excited state. The excited state spectrum is redshifted to lower photon energies with respect to the ground state, which is obvious by the background-free signature at 526.4 eV. This signature must be a new core excitation channel either to the n or π electron hole. An additional signature of UV excitation, the intensity reduction in the area of the π* resonances, is the result of a bleach of the ground state spectrum almost entirely compensated by the redshift of the smooth K-edge ionization feature in the excited state spectrum. The effect is therefore only visible, where the ground state exhibits the strongest intensity modulations, i.e. the π* resonance. It is therefore a direct signature of the ground state depopulation and largely independent of any following excited state dynamics.

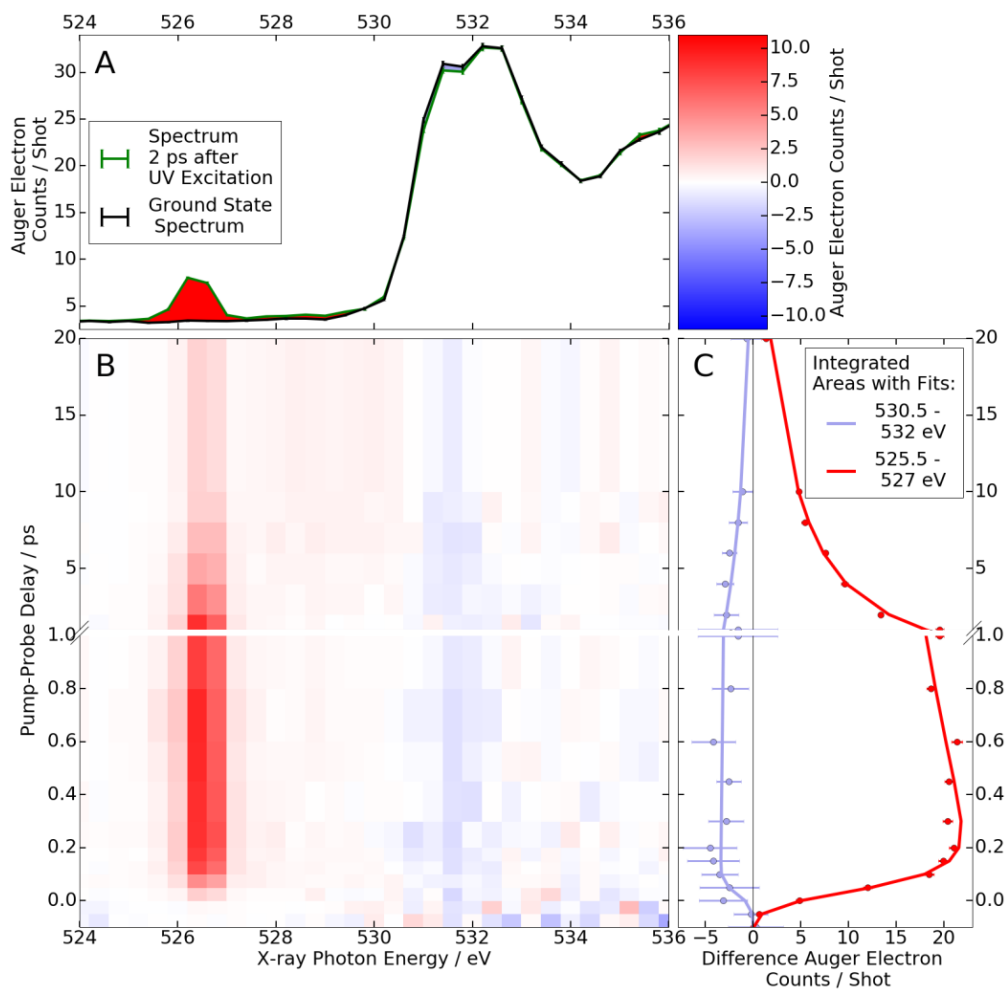

*Fig. 2.* (A) Representative NEXAFS spectra 2ps after UV excitation and without UV excitation. UV-induced increase in intensity is marked red, UV-induced decrease is light blue. UV excitation leads to the appearance of a new spectral feature around 526.4 eV and a bleach of the ground state π* resonance at 531.4 eV. (B) False-color plot of time-dependent NEXAFS difference spectra (see color bar in the upper right corner). The UV-induced features at 526.4 eV and 531.4 eV are clearly visible throughout the positive pump-probe delays. (C) Time-dependence of the UV-induced features with fits based on a rate equation model.

The time-dependence of the difference signal (x-ray absorption with UV minus x-ray absorption without UV) is shown in Fig. 2B. The spectrally integrated time trends of the ground state bleach and excited state features are shown in Fig. 2C. The temporal onset of the excited state feature exhibits a delay ((60 ± 30) fs according to a rate equation fit, see Supporting Material) with respect to the temporal overlap between UV and x-ray pulses, which is marked by the bleach onset. The intensity of the 526.4 eV feature is only due to 13 % of the population in the ground state NEXAFS spectrum. Its absorption cross-section is, thus, similar to the π* resonance. Therefore, it must be likewise due to a localized transition, which is the signature of the nπ* state, not the ππ* state. Accordingly, the delay of the nπ* signature of (60 ± 30) fs directly reflects the nuclear wavepacket dynamics to access the ππ*/nπ* conical intersection seam, in agreement with a short ππ* lifetime observed in our earlier DUXAP study.

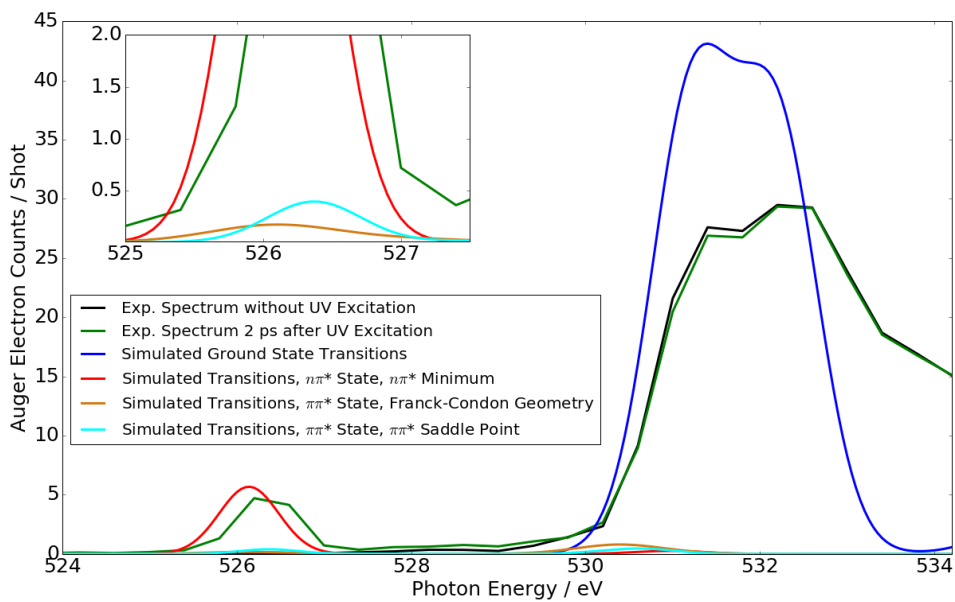

**Fig. 3.** Simulated spectra of the ground state, the ππ* state in the Franck-Condon region and at the saddle point, and the nπ* state at its minimum. For comparison, the experimental ground state spectrum and the experimental spectrum 2 ps after UV excitation are shown. Relative intensities of simulated and experimental ground state spectra are adjusted to an equal peak area of the π* resonances. Simulated excited state spectra are scaled with respect to the ground state spectrum assuming 13 % excitation. The inset shows a detailed view of the intensity relations at the position of the excited state feature. Contributions from K-edge ionization were not included in the simulations.

Our intuitive interpretation is supported by CC NEXAFS spectra simulations of the ground state and of the excited states at the minimum and saddle point geometries identified in Fig. 1B. We compare calculated to experimental spectra of the ground state and 2 ps after UV excitation in Fig. 3. All three simulated excited state spectra exhibit their lowest energy resonance around 526.4 eV. In all cases, the final state is the same O(8)-centered core-excited state (CE). As expected, the oscillator strength in the nπ* state beats the ππ* state by a factor of 40, almost independent of the molecular geometry (see Supporting Material). We scaled the simulated excited state spectra to the estimated ratio of 13 % excited molecules. Only the simulated nπ* state spectrum shows a comparable intensity at the 526.4 eV position.

Thus, with the TR-NEXAFS method presented here, we can confirm population of the nπ* state through a directly accessible conical intersection within 60 fs. Comparison of experimentally observed and calculated NEXAFS absorption intensities suggests that internal conversion into the nπ* state is a major channel in the relaxation dynamics of thymine. The nπ* signature shows a biexponential decay with time constants of (1.9 ± 0.1) ps and (10.5 ± 0.2) ps. This supports a consecutive relaxation process via a level with nπ* character to a final level of non-nπ* character, to which our method is insensitive. The nπ* level is most probably another minimum in the singlet nπ* state, since intersystem crossing to a triplet nπ* state is forbidden by the El Sayed selection rule. The transition to the final state can be either intersystem crossing to a nearby triplet ππ* state with strong spin orbit coupling (*27*) or internal conversion to the ground state (*28*), which is supported by the recovery of the ground state bleach in our data.

In conclusion, we demonstrated with this work a novel method to selectively investigate ultrafast ππ*/nπ* internal conversion. We use the dominating absorption strength of the oxygen 1s-n resonance to directly monitor this nonadiabatic process. An alternative method, however less selectively sensitive to electronic structure changes, is time-resolved photoelectron spectroscopy using extreme ultraviolet pulses. Here one relies on spectrally resolving ionic continua that single out certain valence states (*29*), thus exhibiting nonadiabatic transitions by kinetic energy or angular distribution changes (*30*, *31*). An assignment is, however, only possible by high-level simulations of excited state ionization cross-sections, whereas we confirm the validity of our intuitive orbital-based assignment of core excitations by quantitative CC simulations. Our results prove that the method already works reliably under the current, still challenging conditions of x-ray free-electron laser experiments with low repetition rates and high temporal and spectral jitter. The method has the potential to become a standard tool for ultrafast investigations at the upcoming second generation of ultrafast x-ray sources with up to MHz repetition rates and improved shot-to-shot stability.

**Acknowledgments:**

This work was supported by the AMOS program within the Chemical Sciences, Geosciences, and Biosciences Division of the Office of Basic Energy Sciences, Office of Science, U.S. Department of Energy. Parts of this research were carried out at the Linac Coherent Light Source (LCLS) at the SLAC National Accelerator Laboratory. LCLS is an Office of Science User Facility operated for the U.S. Department of Energy Office of Science by Stanford University. We thank NOTUR for computer time through Project nn2962k. MG acknowledges funding via the Office of Science Early Career Research Program through the Office of Basic Energy Sciences, U.S. Department of Energy and NB under grant No. DE-SC0012376. MG is now funded by a Lichtenberg Professorship from the Volkswagen foundation. TJAW thanks the German National Academy of Sciences Leopoldina for a fellowship (Grant No. LPDS2013-14). HK acknowledges support from FP7-PEOPLE-2013-IOF (Project No. 625321). JG acknowledges the European Research Agency via the FP-7 PEOPLE Program (Marie Curie Action 298210). SC acknowledges support from the AIAS-COFUND program (Grant Agreement No. 609033). RF would like to acknowledge financial support from the Knut and Alice Wallenberg Foundation, Sweden, and the Swedish Research Council.

# Supporting Material for: Probing ultrafast ππ*/*n*π* internal conversion in organic chromophores via K-edge resonant absorption

**Authors:** T. J. A. Wolf[1], R. H. Myhre[1,2], J. P. Cryan[1], S. Coriani[3,4], R. J. Squibb[5], A. Battistoni[1], N. Berrah[6], C. Bostedt[7,8,9], P. Bucksbaum[1,10], G. Coslovich[7], R. Feifel[5], K. J. Gaffney[1], J. Grilj[11], T. J. Martinez[1,12], S. Miyabe[1,12,13], S. P. Moeller[7], M. Mucke[14], A. Natan[1], R. Obaid[6], T. Osipov[7], O. Plekan[15], S. Wang[1], H. Koch[1,2*], M. Gühr[1,16*]

[1]Stanford PULSE Institute, SLAC National Accelerator Laboratory, Menlo Park, CA 94025, USA.
[2]Department of Chemistry, Norwegian University of Science and Technology, NO-7491 Trondheim, Norway.
[3]Dipartimento di Scienze Chimiche e Farmaceutiche, Università degli Studi di Trieste, Italy.
[4]Aarhus Institute of Advanced Studies, Aarhus University, DK-8000 Aarhus C, Denmark.
[5]Department of Physics, University of Gothenburg, Gothenburg, Sweden.
[6]Department of Physics, University of Connecticut, Storrs, Connecticut 06269, USA.
[7]Linac Coherent Light Source, SLAC National Accelerator Laboratory, Menlo Park, California 94720, USA.
[8]Argonne National Laboratory, Lemont, Illinois 60439, USA.
[9]Department of Physics and Astronomy, Northwestern University, Evanston, Illinois 60208, USA.
[10]Department of Physics, Stanford University, Stanford, CA 94305, USA.
[11]Laboratory of Ultrafast Spectroscopy, Ecole Polytechnique Federal de Lausanne, CH 1015 Switzerland.
[12]Department of Chemistry, Stanford University, Stanford, California 94305, USA.
[13]Laser Technology Laboratory, RIKEN, Wako, Saitama 351-0198, Japan
[14]Department of Physics and Astronomy, Uppsala University, Box 516, SE-751 20 Uppsala, Sweden.
[15]Elettra-Sincrotrone Trieste, I-34149 Basovizza, Trieste, Italy.
[16]Institut für Physik und Astronomie, Universität Potsdam, 14476 Potsdam, Germany.

*Correspondence to: mguehr@uni-potsdam.de, henrik.koch@ntnu.no

**Experimental Methods:** The experiment was performed at the LCLS free electron laser facility, SLAC National Accelerator Laboratory, at the soft x-ray (SXR) instrument (*32*, *33*). A schematic representation of the experimental setup is shown in Fig. S1. Thymine was purchased from Sigma Aldrich and evaporated by an effusive oven into an ultra high vacuum chamber at a temperature of 160°C leading to a sample density of $10^{12}$ cm$^{-3}$ in the overlap region of optical and x-ray laser (*23*, *34*). Molecules were excited by 267 nm pulses with 70 fs duration and a focus diameter of 100 µm FWHM. Soft x-ray pulses with 70 fs duration and a focus diameter of 70 µm FWHM were used to probe the sample in the oxygen K-edge spectral region from 520 to 550 eV by simultaneously tuning the FEL and the monochromator of the SXR instrument with an energy resolution of < 0.5 eV (*35*). The intensity of the essentially background-free transient feature at 526.4 eV was measured for a wide range of UV pump intensities, to make sure the experiment took place in the linear absorption regime well below saturation. Temporal and spatial overlap of UV and SXR pulses was optimized to a bleach in the Auger spectra of thymine induced by photofragmentation at high UV intensities. Oxygen 1s Auger spectra were recorded with the 2m long LCLS-FELCO magnetic bottle spectrometer (*36*). The photon energy dependent absorption cross-section of the sample is proportional to the integrated Auger electron yield. SXR pulses were delayed with respect to UV pulses between -200 fs and 20 ps. To achieve NEXAFS difference spectra, UV laser pulses were blocked on a shot-by-shot basis. LCLS pulses are strongly fluctuating in intensity and relative arrival time. Therefore, both parameters were recorded on a shot-by-shot basis by an optical x-ray cross-correlator (*37*) and a gas detector after the monochromator, respectively. The dataset was resorted into ≥ 50 fs delay bins and several x-ray intensity bins. Difference spectra from different x-ray intensity bins were averaged

**Theoretical Methods:** The thymine ground state geometry (Tab. S1) was optimized with CCSD(T)/aug-cc-pVDZ using CFOUR (*38*). Excited state geometries (Tab. S2 and S3) were optimized at the EOM-CCSD/aug-cc-pVDZ level employing Q-Chem (*39*). No symmetry restrictions were applied for geometry optimizations. Valence excitation energies were obtained with CC3 using the aug-cc-pCVDZ basis on the oxygens and the aug-cc-pVDZ basis on the other atoms. We employed a newly developed implementation in Dalton (*16*, *40–42*). Oxygen 1s to valence excitation energies and oscillator strengths were computed at the CCSD level of theory with the same basis as for the valence excitations using a newly developed linear response code employing core-valence separation and implemented in Dalton (*18*, *43–45*) (see Tabs. S7-S9). The core to valence excitation energies were offset-corrected by benchmark calculations of the lowest core to valence excitation energies at the CC3/aug-cc-pCVTZ/aug-cc-pVDZ level. With this correction we achieve quantitative agreement with the NEXAFS transition energies within the experimental error bars on a purely ab initio basis. The theoretical core excitation energies are not corrected for relativistic effects and we estimate the effect to increase excitation energies by 0.1-0.3 eV.

Thymine exhibits $C_s$ symmetry in the ground state. The two lowest-lying excited states have different representations, A´´(nπ*) and A´(ππ*), and the ππ*/nπ* conical intersection is symmetry allowed. We note that no complex eigenvalues of the Jacobian matrix were encountered in the vicinity of the conical intersection seam (*46*, *47*). In contrast to earlier studies (*20*, *21*, *48–50*), we could not identify a minimum in the ππ* state. Instead we found a saddle point geometry with $C_s$ symmetry, which is directly accessible from the Franck-Condon region. The energy lowering degrees of freedom of the saddle point are out of plane bending, as confirmed by frequency calculations, which were performed for all observed stationary points (*51*, *52*) (Tabs. S4-S6). The nπ* minimum geometry is distorted from $C_s$ symmetry with O(8) out of the plane. We encountered the ππ*/nπ* conical intersection seam in between the Franck-Condon point and the ππ* saddle point in close proximity to the latter (energy difference < 0.03 eV). Based on calculated core excitation energies and oscillator strengths, NEXAFS spectra were simulated by convoluting theoretical stick spectra with Gaussians to account for peak broadening and the experimental energy resolution.

**Localization of excitations and the single electron picture:** For our intuitive prediction of the expected intensities of excited state NEXAFS features we rely on the Hartree-Fock (HF) molecular orbital (MO) based one-electron picture. In this framework, core-excited as well as valence-excited states are described as a single electron excitation from an occupied to an unoccupied HF-MO. Despite its success in the case of relative intensities for core-excitations from the $\pi\pi^*$ and $n\pi^*$ states, it fails in predicting relative intensities for core-excitations from the ground state. The ground state core-excitation leads to the $\pi^*$ MO in the one-electron picture, which is highly delocalized and exhibits only weak density at the oxygens. Nevertheless, both coupled cluster simulations and experimental results predict the transition intensity to be comparable to the $n\pi^*$ state core-excitation, which involves the strongly localized n MO.

The reason for the failure of the MO picture lies in the HF formalism, which only optimizes occupied MOs. The employed coupled cluster methods calculate valence and core excited states by optimizing a series expansion of electron excitations into the virtual MOs inherently leaving the single electron MO picture. The most appropriate, but less intuitive way to inspect these excitations is therefore to look at electron density changes instead of orbitals (see Fig. S2). The density changes, however, qualitatively agree with the MO picture in cases of transitions between MOs which are occupied in the HF reference wavefunction.

**Rate equation model:** To analyze the transient 526.4 eV feature in the TR-NEXAFS spectra, a rate equation model was employed, which assumes the following chain of subsequent excited state single exponential population transfers with time constants $\tau_1$ to $\tau_3$:

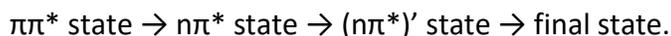

$\pi\pi^*$ state → $n\pi^*$ state → $(n\pi^*)'$ state → final state.

Since the 526.4 eV signature decays in a bi-exponential fashion, the step labeled as $(n\pi^*)'$ had to be included. The character of that step cannot be completely determined by the present results. Its transition moment is only 42 % of the $n\pi^*$ state according to our fit of the experimental data. It is nevertheless very likely also of $n\pi^*$ character, since the calculated oscillator strength for core excitation from the $\pi\pi^*$ state is only 2.5 % of the oscillator strength from the $n\pi^*$ state and, thus, one order of magnitude lower. One possible explanation is that thymine relaxes thermally out of the $n\pi^*$ minimum identified in our calculations into a lower lying minimum, from which it then relaxes to the final state, which is either the ground state or a triplet state.

Since the bleach signature only decays slowly within the investigated delay-time window, it is ideal to extract the exact time zero and the instrument response function (90 fs) using an error function fit. To decrease the noise level, a region of interest with the strongest UV-induced modulations was identified in the resonant Auger spectra from the ground state $\pi^*$ resonances. The plotted intensities in Fig. S3 refer to the integrated UV-induced changes in this region of interest. For comparison, Fig. S3 also shows the analogue signal for the 526.4 eV, which was fitted (see Fig. 2 and Fig. S3) with a weighted sum of the time-dependent populations of the nπ* and (nπ*)' steps, convoluted with the instrument response function (g(t)).

$$I(\Delta t) = g(t) \otimes heaviside(\Delta t)$$
$$\cdot \left( I_1 \frac{\tau_2}{\tau_2 - \tau_1} \left( e^{-\frac{\Delta t}{\tau_2}} - e^{-\frac{\Delta t}{\tau_1}} \right) \right.$$
$$\left. + I_2 \frac{\tau_3}{\tau_2 - \tau_1} \left( \frac{\tau_2}{\tau_3 - \tau_2} \left( e^{-\frac{\Delta t}{\tau_3}} - e^{-\frac{\Delta t}{\tau_2}} \right) - \frac{\tau_1}{\tau_3 - \tau_1} \left( e^{-\frac{\Delta t}{\tau_3}} - e^{-\frac{\Delta t}{\tau_1}} \right) \right) \right)$$

The delay between the onsets of the bleach and the 526.4 eV feature is clearly visible in Fig. S3.

**Excited state population analysis**: Calculated and experimental intensities of the nπ* feature and the ground state π* resonance have to be compared, to estimate, which fraction of the excited state population is observable in the nπ* feature. For this, we chose the same NEXAFS spectra of the ground state and 2 ps after UV excitation as in Fig. 2A. The ratio of integrated peak areas between the nπ* ($I_2$) and ground state resonance features ($I_1$) is 0.053. These can now be compared to the ratio of calculated transition moments for the nπ* feature ($\sigma_2$) and the ground state resonance ($\sigma_1$) of 0.65. Assuming 100 % population transfer from the ππ* state, the rate equation model predicts the nπ* level to contain 36 % and the (nπ*)' level 57 % of the overall relative excited state population ($P_0$). According to the rate equation fit, the transition moment of (nπ*)' level $\sigma_2'$ is 42 % of the nπ* transition moment $\sigma_2$. This leads to the relation for $I_2$

$$I_2 = f \cdot P_0 \cdot \sigma_2 \cdot (0.36 + 0.57 \cdot 0.42)$$

where f is a conversion factor between experimental intensities and calculated transition moments. The factor f can be evaluated by $f = \frac{I_1}{\sigma_1}$. Inserting this in the equation for $I_2$ gives a value of 13 % for the overall excited state population $P_0$ relative to the ground state. This moderate excitation ratio fits well together with our expectations of the excitation ratio based on our scan of the UV intensity dependence of the nπ* feature intensity. Comparison with the intensity dependence furthermore supports the initial assumption that the relaxation to the nπ* state observed in the present experiment is a major channel for the ππ* population.

**Intensity scan:** To confirm that UV excitation takes place in the linear regime, we investigated the dependence of the nπ* feature intensity on the UV intensity (see Fig. S4). At low UV intensities, the nπ* feature intensity has a linear response. At high intensities, saturation is observable. The most relevant processes contributing to the excited state population are single photon excitation from the ground state and further single photon excitation from the excited state i.e. sequential or resonance-enhanced two photon excitation. The dependence of the relative population $P_0$ in the excited state on the photon flux F is therefore

$$P_0 = \frac{\sigma_1}{\sigma_1 - \sigma_2}[e^{-\sigma_2 \cdot F} - e^{-\sigma_1 \cdot F}]$$

where $\sigma_1$ and $\sigma_2$ are the ground state and excited state absorption cross-sections. The value of $\sigma_1$ is approximately 30 MBarn, the value of $\sigma_2$ is unknown. Since absorption of two photons brings the molecule very close to the ionization threshold where the density of states is particularly high, the value of $\sigma_2$ can be expected to be higher than $\sigma_1$. The absolute upper limit for relative excited state population can be estimated by neglecting sequential two photon excitations i.e. by setting $\sigma_2$ to zero. In this case, the saturation value refers to 100 % population in the excited state. The UV intensity for the pump-probe experiments leads to 24 % of this saturation value. Assuming any value higher than 0 for $\sigma_2$ reduces the population in the excited state at saturation. Assuming 13 % excitation at the intensity of the pump probe experiments based on the comparison between experimental and calculated intensities yields a value for $\sigma_2$, which is 1.5 times $\sigma_1$ and thus perfectly reasonable.

### Figures S1-S4:

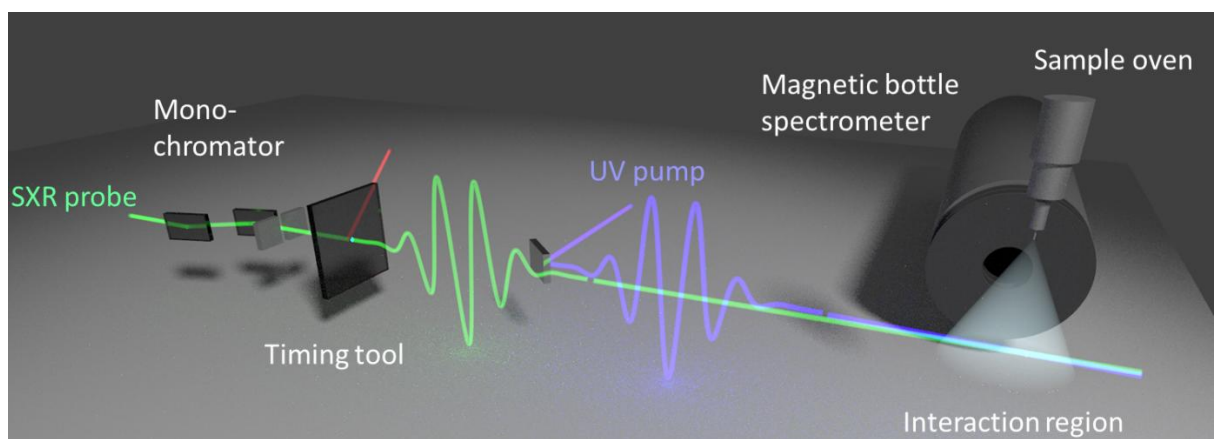

**Fig. S1.** Schematic representation of the experimental setup. Broad-bandwidth ultrashort soft x-ray (SXR) pulses (green) are monochromatized and focused into the interaction region of the experimental chamber. There, they are quasi-collinearly overlapped with ultrafast UV pulses with a wavelength of 267 nm (violet). The sample thymine is evaporated into the interaction region by an in-vacuum oven. The relative timing between UV and soft x-ray pulses is measured on a shot-by-shot basis using the SXR timing tool to compensate for the 200 fs timing jitter. Auger electron spectra from core-excited thymine molecules are detected in a magnetic bottle photoelectron spectrometer. NEXAFS spectra can be generated by measuring the photon energy dependent integrated Auger electron yield.

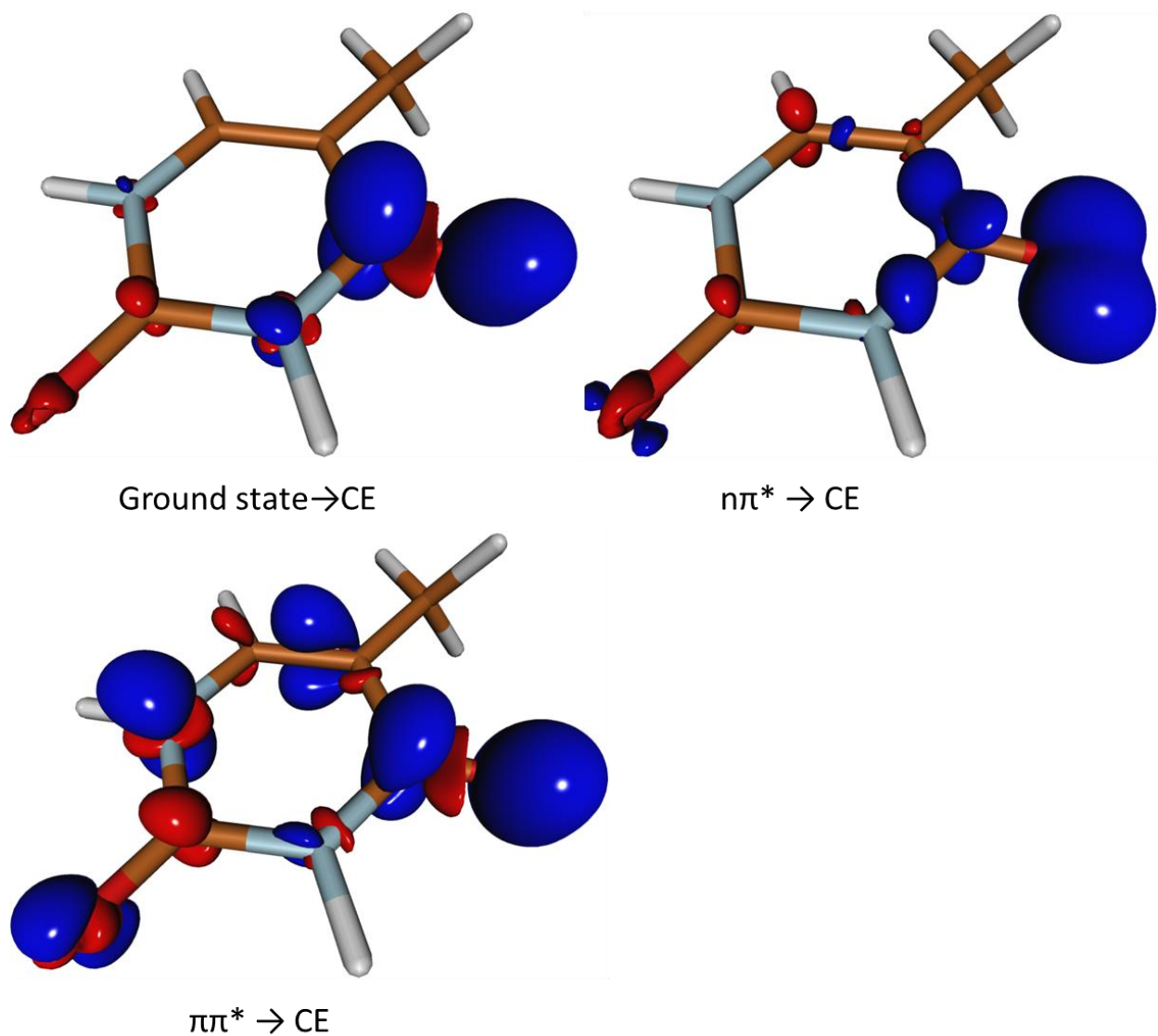

**Fig. S2.** Visualization of electron density changes for core excitations from the ground state, the nπ* state and the ππ* state to the lowest O(8) 1s core-excited state. In the case of core excitation from the nπ* state and the ππ* state, the electron density changes agree with the predictions from the Hartree-Fock (HF) molecular orbital (MO) based single electron picture. The density change for the nπ* state core excitation is strongly localized at O(8), since two MOs with strong localization at this oxygen are involved. The density change for the ππ* state core excitation is delocalized, since it involves delocalized π MO apart from the localized O(8) 1s MO. The electron density change for the core excitation from the ground state disagrees with the predictions from the HF MO picture. The latter involves the localized O(8) 1s MO and a delocalized π* MO. The electron density change, however, is strongly localized due to linear combination of single electron transitions to several π* MOs.

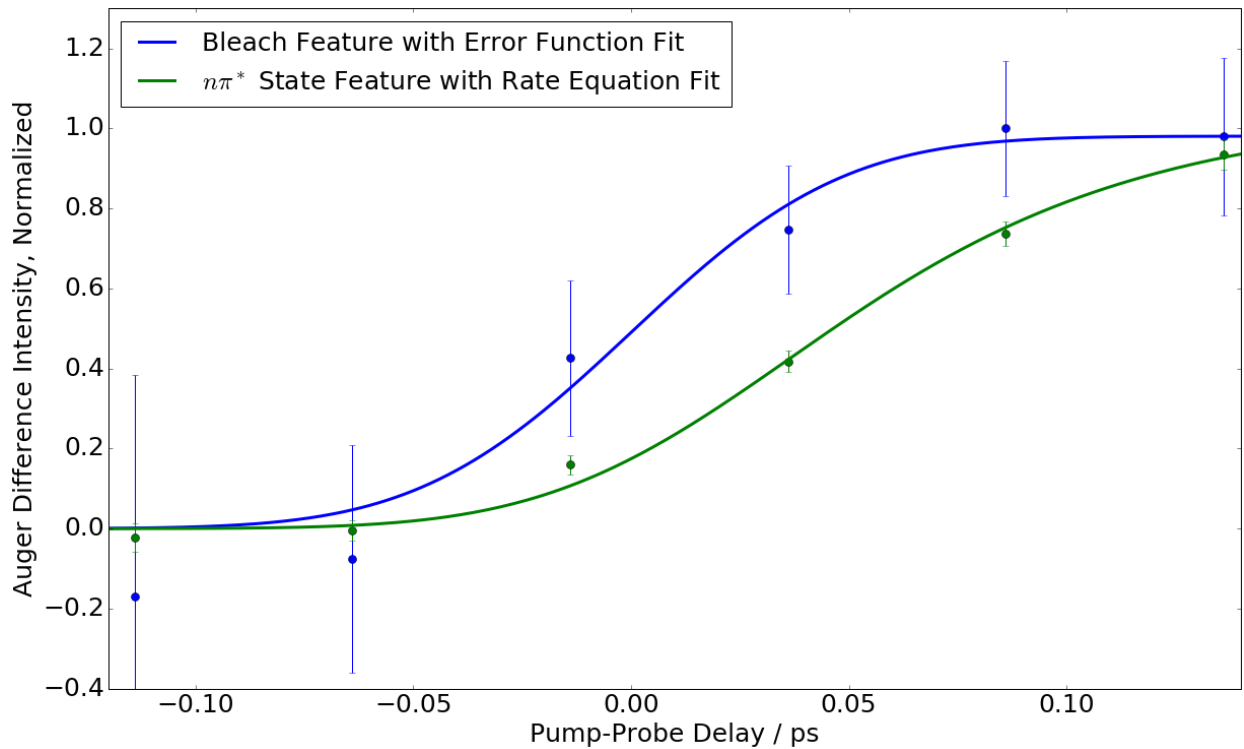

**Fig. S3:** Representation of the fit of the bleach signal (blue) used to determine the experimental response function and time zero. The experimental data are extracted from a region of interest in the Auger spectra from the photon energy region, where the bleach is observable in the NEXAFS spectra. For better comparison with the corresponding dataset from the region of the nπ* feature (green), the bleach dataset is inverted. Both datasets are normalized to the maximum modulation. The time-dependence of nπ* feature is fitted with the rate equation model. The delay between the bleach and the nπ* feature is clearly visible.

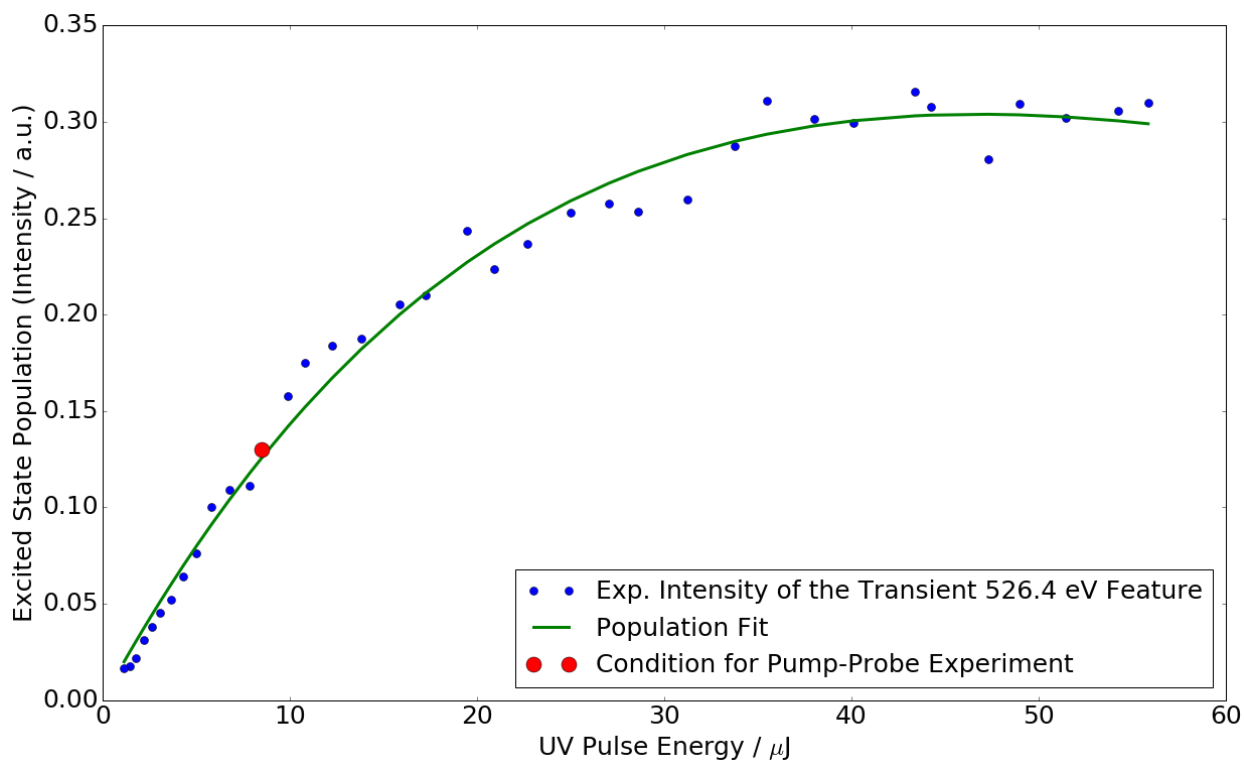

Fig. S4: Intensity of the transient 526.4 eV feature for different UV intensities together with a rate equation fit. The ordinate is rescaled to match the estimated 13 % excited state population for the UV intensity used in the pump–probe experiment.

**Tables S1-S12:**

**Table S1: Ground state minimum geometry (CCSD(T)/aug-cc-pVDZ)**

| Element | X / Å | Y / Å | Z / Å |
|---|---|---|---|
| C | -1.6332 | 0.0426 | -0.0001 |
| C | 0.7570 | 0.8384 | -0.0001 |
| C | 1.1979 | -0.5713 | 0.0000 |
| C | 0.2400 | -1.5432 | 0.0000 |
| C | 2.6817 | -0.8510 | 0.0002 |
| N | -0.6415 | 1.0215 | -0.0007 |
| N | -1.1219 | -1.2509 | -0.0002 |
| O | -2.8353 | 0.2919 | 0.0005 |
| O | 1.5118 | 1.8105 | 0.0003 |
| H | -0.9734 | 1.9835 | -0.0002 |
| H | 0.4869 | -2.6087 | 0.0000 |
| H | -1.8153 | -1.9901 | 0.0005 |

| Element | X / Å | Y / Å | Z / Å |
|---|---|---|---|
| H | 3.1589 | -0.4028 | -0.8884 |
| H | 3.1587 | -0.4025 | 0.8887 |
| H | 2.8721 | -1.9376 | 0.0005 |

Table S2: nπ* state minimum geometry (EOM-CCSD/aug-cc-pVDZ)

| Element | X / Å | Y / Å | Z / Å |
|---|---|---|---|
| C | 1.6756 | 0.0108 | 0.0230 |
| C | -0.6393 | 0.8270 | -0.0428 |
| C | -1.1709 | -0.4819 | -0.0168 |
| C | -0.2637 | -1.5185 | 0.0106 |
| C | -2.6680 | -0.6951 | -0.0181 |
| N | 0.7453 | 1.0515 | 0.1200 |
| N | 1.1227 | -1.2492 | 0.0249 |
| O | 2.8776 | 0.2283 | -0.0350 |
| O | -1.3720 | 1.9491 | 0.1211 |
| H | 1.1236 | 1.9641 | -0.1066 |
| H | -0.5521 | -2.5678 | -0.0154 |
| H | 1.7941 | -2.0025 | -0.0250 |
| H | -3.1289 | -0.2542 | 0.8817 |
| H | -3.1341 | -0.2281 | -0.9015 |
| H | -2.8981 | -1.7710 | -0.0329 |

Table S3: ππ* state saddle point geometry (EOM-CCSD/aug-cc-pVDZ)

| Element | X / Å | Y / Å | Z / Å |
|---|---|---|---|
| N | 0.6451 | -1.0681 | -0.0013 |
| H | 0.9407 | -2.0373 | -0.0009 |
| C | 1.6033 | -0.1081 | 0.0001 |
| O | 2.8197 | -0.2668 | 0.0006 |
| N | 1.1090 | 1.2319 | 0.0005 |
| H | 1.8635 | 1.9158 | 0.0010 |
| C | -0.1892 | 1.6224 | -0.0010 |
| H | -0.4105 | 2.6884 | -0.0014 |
| C | -0.7824 | -0.8153 | -0.0001 |
| O | -1.5291 | -1.8220 | 0.0005 |
| C | -1.1881 | 0.5501 | -0.0002 |
| C | -2.6474 | 0.8865 | 0.0006 |
| H | -2.8068 | 1.9771 | -0.0001 |
| H | -3.1465 | 0.4513 | 0.8852 |

| H | -3.1475 | 0.4502 | -0.8830 |

Table S4: Vibrational frequencies and IR intensities of the ground state minimum (CCSD(T)/aug-cc-pVDZ)

| Frequency / cm$^{-1}$ | IR intensity / km mol$^{-1}$ |
|---:|---:|
| 100 | 0.00 |
| 134 | 0.72 |
| 143 | 0.01 |
| 269 | 2.41 |
| 279 | 0.12 |
| 376 | 22.94 |
| 381 | 21.25 |
| 451 | 18.44 |
| 525 | 52.61 |
| 536 | 6.68 |
| 594 | 1.37 |
| 665 | 88.91 |
| 730 | 4.42 |
| 735 | 25.94 |
| 754 | 1.88 |
| 797 | 3.42 |
| 875 | 19.64 |
| 960 | 11.63 |
| 1011 | 1.94 |
| 1052 | 0.12 |
| 1154 | 9.39 |
| 1199 | 154.13 |
| 1249 | 3.83 |
| 1373 | 12.10 |
| 1386 | 0.06 |
| 1403 | 6.08 |
| 1424 | 66.38 |
| 1453 | 6.06 |
| 1477 | 1.66 |
| 1499 | 111.43 |
| 1699 | 0.03 |
| 1739 | 555.62 |
| 1778 | 817.99 |
| 3034 | 23.89 |
| 3109 | 10.51 |
| 3123 | 14.47 |
| 3212 | 4.81 |
| 3594 | 60.42 |
| 3643 | 96.73 |

Table S5: Vibrational frequencies and IR intensities of the nπ* state minimum (EOM-CCSD/aug-cc-pVDZ)

| Frequency / cm$^{-1}$ | IR intensity / km mol$^{-1}$ |
|---:|---:|
| 82 | 2.64 |
| 100 | 1.89 |
| 135 | 0.51 |
| 181 | 19.25 |
| 247 | 4.70 |
| 262 | 2.22 |
| 337 | 4.95 |
| 406 | 30.45 |
| 460 | 21.14 |
| 468 | 91.00 |
| 510 | 15.10 |
| 521 | 47.83 |
| 576 | 6.25 |
| 637 | 47.10 |
| 741 | 33.82 |
| 753 | 3.86 |
| 794 | 7.54 |
| 942 | 26.29 |
| 1012 | 1.58 |
| 1060 | 0.32 |
| 1146 | 62.73 |
| 1211 | 44.59 |
| 1243 | 13.73 |
| 1283 | 5.37 |
| 1390 | 84.14 |
| 1419 | 0.70 |
| 1432 | 13.49 |
| 1476 | 22.63 |
| 1476 | 6.40 |
| 1492 | 0.86 |
| 1516 | 27.52 |
| 1627 | 45.54 |
| 1800 | 742.28 |

| Frequency / cm$^{-1}$ | IR intensity / km mol$^{-1}$ |
|---|---|
| 3048 | 24.42 |
| 3115 | 13.00 |
| 3144 | 13.12 |
| 3265 | 1.64 |
| 3637 | 70.62 |
| 3680 | 76.92 |

Table S6: Vibrational frequencies and IR intensities of the ππ* state saddle point (EOM-CCSD/aug-cc-pVDZ)

| Frequency / cm$^{-1}$ | IR intensity / km mol$^{-1}$ |
|---|---|
| 292i | 31.79 |
| 126i | 0.25 |
| 94i | 13.22 |
| 129 | 0.54 |
| 153 | 0.30 |
| 271 | 7.96 |
| 300 | 5.65 |
| 372 | 18.71 |
| 440 | 35.66 |
| 441 | 76.76 |
| 496 | 11.09 |
| 578 | 0.36 |
| 610 | 0.16 |
| 662 | 4.14 |
| 726 | 12.06 |
| 754 | 65.88 |
| 768 | 9.75 |
| 925 | 17.76 |
| 968 | 4.02 |
| 996 | 0.39 |
| 1150 | 59.95 |
| 1185 | 30.92 |
| 1254 | 12.70 |
| 1319 | 32.57 |
| 1351 | 11.99 |
| 1387 | 90.04 |
| 1412 | 41.87 |
| 1443 | 7.86 |
| 1466 | 22.10 |
| 1480 | 36.99 |
| 1524 | 1.79 |
| 1620 | 263.15 |
| 1762 | 320.77 |
| 3022 | 14.56 |
| 3080 | 11.34 |
| 3121 | 14.83 |
| 3275 | 5.94 |
| 3588 | 68.66 |
| 3650 | 99.19 |

Table S7: Calculated NEXAFS resonance energies and cross-sections for different states at the Franck-Condon geometry (CCSD/aug-cc-pCVDZ/aug-cc-pVDZ)

| O(8) transitions / eV | Oscillator strength | O(7) transitions / eV | Oscillator strength |
|---:|---:|---:|---:|
| Ground state | | | |
| 535.46 | 3.41E-02 | 536.45 | 3.17E-02 |
| 538.95 | 3.60E-04 | 538.75 | 1.34E-04 |
| 539.86 | 1.51E-03 | 539.55 | 2.39E-03 |
| 539.95 | 2.30E-03 | 539.70 | 3.69E-04 |
| 539.97 | 8.26E-04 | 540.10 | 1.53E-03 |
| 540.27 | 8.53E-05 | 540.21 | 1.01E-03 |
| 540.54 | 1.74E-03 | 540.42 | 2.99E-03 |
| 540.79 | 1.11E-04 | 540.93 | 5.19E-04 |
| 541.05 | 1.01E-04 | 541.14 | 1.00E-03 |
| 541.14 | 6.98E-05 | 541.44 | 7.24E-04 |
| nπ* state | | | |
| 530.56 | 3.68E-02 | 531.55 | 1.10E-06 |
| 534.05 | 8.90E-07 | 533.85 | 1.00E-08 |
| 534.96 | 1.20E-06 | 534.65 | 1.30E-07 |
| 535.05 | 1.18E-03 | 534.80 | 1.20E-03 |
| 535.07 | 1.80E-07 | 535.20 | 4.75E-05 |
| 535.37 | 6.15E-04 | 535.31 | 2.20E-07 |
| 535.64 | 7.00E-08 | 535.52 | 1.00E-08 |
| 535.89 | 1.00E-08 | 536.03 | 2.00E-08 |
| 536.15 | 1.40E-07 | 536.24 | 1.00E-08 |
| 536.24 | 1.00E-08 | 536.54 | 1.27E-05 |
| ππ* state | | | |
| 530.33 | 1.74E-03 | 531.32 | 3.04E-04 |
| 533.82 | 1.09E-04 | 533.62 | 7.22E-05 |
| 534.73 | 3.80E-05 | 534.42 | 2.32E-05 |
| 534.82 | 1.17E-03 | 534.57 | 6.46E-03 |
| 534.84 | 5.29E-05 | 534.97 | 1.61E-04 |
| 535.14 | 9.83E-04 | 535.08 | 1.31E-06 |

| O(8) transitions / eV | Oscillator strength | O(7) transitions / eV | Oscillator strength |
|---|---|---|---|
| 535.41 | 7.09E-06 | 535.29 | 4.00E-07 |
| 535.66 | 2.90E-07 | 535.80 | 2.10E-07 |
| 535.92 | 1.00E-08 | 536.01 | 2.12E-06 |
| 536.01 | 2.33E-06 | 536.31 | 4.43E-05 |

Table S8: Calculated NEXAFS resonance energies and cross-sections for different states at the nπ* minimum geometry (CCSD/aug-cc-pCVDZ/aug-cc-pVDZ)

| O(8) transitions / eV | Oscillator strength | O(7) transitions / eV | Oscillator strength |
|---|---|---|---|
| Ground state | | | |
| 534.49 | 3.14E-02 | 536.39 | 3.12E-02 |
| 538.53 | 2.55E-04 | 538.71 | 5.42E-04 |
| 539.31 | 4.45E-04 | 539.05 | 2.16E-03 |
| 539.38 | 2.07E-03 | 539.55 | 3.57E-03 |
| 539.45 | 1.69E-03 | 540.06 | 2.89E-03 |
| nπ* state | | | |
| 530.80 | 4.30E-02 | 532.70 | 1.31E-05 |
| 534.84 | 5.10E-07 | 535.02 | 1.00E-07 |
| 535.62 | 3.00E-08 | 535.36 | 8.43E-04 |
| 535.69 | 6.10E-07 | 535.86 | 0.00E+00 |
| 535.76 | 1.43E-03 | 536.37 | 5.00E-06 |
| ππ* state | | | |
| 529.88 | 7.67E-03 | 531.78 | 2.84E-04 |
| 533.92 | 9.50E-05 | 534.10 | 5.13E-05 |
| 534.70 | 9.73E-05 | 534.44 | 3.57E-03 |
| 534.77 | 2.60E-04 | 534.94 | 1.26E-05 |
| 534.84 | 3.56E-03 | 535.45 | 1.54E-05 |

Table S9: Calculated NEXAFS resonance energies and cross-sections for different states at the ππ* saddle point geometry (CCSD/aug-cc-pCVDZ/aug-cc-pVDZ)

| O(8) transitions / eV | Oscillator strength | O(7) transitions / eV | Oscillator strength |
|---|---|---|---|
| Ground state | | | |
| 535.28 | 3.25E-02 | 536.42 | 3.11E-02 |
| 538.82 | 2.55E-04 | 538.72 | 7.51E-05 |
| 539.64 | 3.15E-03 | 539.21 | 5.46E-04 |
| 539.67 | 1.32E-03 | 539.51 | 2.62E-03 |
| 539.76 | 1.48E-03 | 540.07 | 1.69E-03 |
| nπ* state | | | |
| 530.88 | 3.91E-02 | 532.02 | 4.32E-04 |
| 534.42 | 4.90E-07 | 534.32 | 6.27E-05 |
| 535.24 | 2.56E-03 | 534.81 | 4.80E-03 |

| 535.27 | 3.66E-06 | 535.11 | 1.38E-05 |
| 535.36 | 8.70E-07 | 535.67 | 7.85E-06 |
| ππ* state | | | |
| 530.99 | 2.98E-03 | 532.13 | 1.11E-04 |
| 534.53 | 6.63E-05 | 534.43 | 3.30E-07 |
| 535.35 | 2.56E-03 | 534.92 | 1.45E-03 |
| 535.38 | 4.53E-05 | 535.22 | 6.00E-08 |
| 535.47 | 7.94E-05 | 535.78 | 2.42E-06 |

**Table S10: Calculated ground state NEXAFS resonance energies at different geometries (CC3/aug-cc-pCVTZ/aug-cc-pVDZ)**

|  | Franck-Condon geometry | nπ* minimum geometry | ππ* saddle point geometry |
| --- | --- | --- | --- |
| Lowest O(8) transition / eV | 531.19 | 529.83 | 530.62 |
| Lowest O(7) transition / eV | 532.23 | 531.86 | 531.84 |

**Table S11: Calculated four lowest valence excitation energies at different geometries (CC3/aug-cc-pCVDZ/aug-cc-pVDZ). The electronic character of the excitation is included in brackets.**

|  | Franck-Condon geometry | nπ* minimum geometry | ππ* saddle point geometry |
| --- | --- | --- | --- |
| $S_1$ / eV | 4.90 (nπ*) | 3.69 (nπ*) | 4.29 (ππ*) |
| $S_2$ / eV | 5.13 (ππ*) | 4.61 (ππ*) | 4.40 (nπ*) |
| $S_3$ / eV | 5.65 (πn*) | 5.12 (ππ*) | 5.26 (πn*) |
| $S_4$ / eV | 6.17 (ππ*) | 5.70 (πn*) | 5.60 (ππ*) |